\documentclass[draft,aps,twocolumn,showpacs,floats,
superscriptaddress]{revtex4}  
\usepackage{amsmath}
\usepackage{amssymb}
\usepackage{graphics,psfig}

\newcommand{\gsim}{ \mathop{}_{\textstyle \sim}^{\textstyle >} }
\newcommand{\lsim}{ \mathop{}_{\textstyle \sim}^{\textstyle <} }
\begin{document}

\preprint{ICRR-Report-497-2003-??}
\preprint{YITP-03-??}
\title{Explosive Dark Matter Annihilation}
\author{Junji Hisano,$^1$ Shigeki Matsumoto,} 
\affiliation{ICRR, University of Tokyo,
Kashiwa 277-8582, Japan }
\author{and Mihoko M. Nojiri}
\affiliation{YITP, Kyoto University, Kyoto 606-8502, Japan}
\date{\today}

\begin{abstract}

If the Dark Matter (DM) in the Universe has interactions with the
standard-model particle, the pair annihilation may give the imprints
in the cosmic ray. In this paper we study the pair annihilation
processes of the DM, which is neutral, however has the electroweak
(EW) gauge non-singlet.  In this estimation the non-relativistic (NR)
effective theory in the EW sector is a suitable technique. We find
that if the DM mass is larger than about 1~TeV, the attractive Yukawa
potentials induced by the EW gauge bosons have significant effects on
the DM annihilation processes, and the cross sections may be enhanced
by several orders of magnitude, due to the zero energy resonance under
the potentials.  Especially, the annihilation to two $\gamma$'s might
have a comparable cross section to other tree-level processes, while
the cross section under the conventional calculation is suppressed by
a loop factor. We also discuss future sensitivities to the $\gamma$
ray from the galactic center by the GLAST satellite detector and the
Air Cerenkov Telescope (ACT) arrays.

\end{abstract}


\maketitle

Nature of the Dark Matter (DM) in the Universe is an important problem
in both particle physics and cosmology.  The Weakly-Interacting
Massive Particle (WIMP), $\chi^0$, is a good candidate for the DM.  It
works as the cold dark matter in the structure formation in the
Universe. High resolution $N$-body simulations show that the cold dark
matter hypothesis explains well the structure larger than about 1 Mpc
\cite{nbody}. Also, the WMAP measured the cosmological abundance
precisely as $\Omega_{DM}=0.27\pm0.04$
\cite{wmap}. Now we know the gravitational property of the DM
in the structure formation and the abundance and distribution in the
cosmological scale. The next questions are the constituent of the DM and
the distribution in the galactic scale.

If the DM is SU(2)$_L$ non-singlet, a pair of the DM could annihilate into
the standard-model (SM) particles with significant cross sections
\cite{DMCR}.  We call such DM's as electroweak-interacting massive
particle (EWIMP) DM in this paper. The detection of exotic cosmic ray
fluxes, such as positron, anti-proton and $\gamma$ ray, may be a
feasible technique to search for the DM's.  Since some DM candidates in
the supersymmetric (SUSY) models have interactions with the SM
particles, these annihilation processes are extensively studied.
Especially, excess of monochromatic $\gamma$ ray due to the pair
annihilation is a robust signal if observed, because the diffused
$\gamma$-ray background must have a continuous energy spectrum
\cite{Bergstrom:1997fj}.  Searches for the exotic $\gamma$ ray from
the galactic center, the galactic halo, or even from extra galaxies are
ones of the projects in the GLAST satellite detector and the big Air
Cerenkov Telescope (ACT) arrays such as CANGAROO III, HESS, MAGIC and
VERITAS. 

In the previous estimates, the cross sections for the EWIMP are
evaluated at the leading order in the perturbation. However, the DM is
non-relativistic (NR) in the current Universe. In this case, if the
EWIMP mass $m$ is much heavier than EW scale, the EWIMP wave function
may be deformed under the Yukawa potentials induced by the EW gauge
boson exchanges and it may give a non-negligible effect in the
annihilation processes.  Furthermore, the neutral EWIMP should has a
charged SU(2)$_L$ partner, $\chi^{\pm}$. When the EWIMP is heavier
than EW scale, their masses are almost degenerate, and the
unsuppressed transition between the two-body states of $2\chi^0$
and $\chi^-\chi^+$ may play an important role in the $2\chi^0$
pair annihilation.

In this letter we reevaluate the pair annihilation cross sections of
the EWIMP's, for the two cases that the DM is a component of two
SU(2)$_L$-doublet fermions or of an SU(2)$_L$-triplet fermion.  These
correspond to the Higgsino-like and Wino-like DM's in the SUSY models,
respectively.  Most interesting fact we find is that the
annihilation cross sections to the gauge boson pairs for the
SU(2)$_L$-doublet (triplet) DM suffer from a zero energy resonance
around $m\simeq 6(2)$~TeV, whose binding energy is zero \cite{landau}
under the potential. Therefore, the cross sections would be enhanced
significantly compared with ones in the perturbative estimations for
$m\gsim 1 (0.5)$~TeV. Furthermore, it is found that the cross section
for $2\chi^0\rightarrow 2\gamma$, which is usually suppressed by a
one-loop factor, becomes comparable to the other tree-level processes,
such as $2\chi^0\rightarrow W^+W^-$, around the resonance. This means
that the mixing between the two-body states of $\chi^-\chi^+$ and $2
\chi^0$ is maximal under the potential.  Due to the explosive
enhancement of the cross sections, the SU(2)$_L$-triplet DM is already
partially constrained by the EGRET observation of the $\gamma$ ray
from the galactic center, and the future $\gamma$ ray searches may
have sensitivity to the heavier EWIMP DM.

First, we summarize properties of the EWIMP DM's. If the DM has a
vector coupling to the $Z$ boson, the current bound from the direct DM
searches through the spin-independent interaction \cite{ds} is
stringent.  This means that the EWIMP DM should be a Majorana fermion
or a real scalar if it is relatively light.  Here we consider a former
case for simplicity.

A simple example for the EWIMP DM's is a neutral component of an
SU(2)$_L$-triplet fermion ($T$) whose hypercharge is zero. This
corresponds to the Wino-like LSP  
in the SUSY models. It is accompanied with the a
charged fermion, $\chi^{\pm}$. While $\chi^0$ and $\chi^{\pm}$ are
almost degenerate in mass in the SU(2)$_L$ symmetric limit, the EW
symmetry breaking by the Higgs field, $h$, generates the mass
splitting, $\delta m$.  If $\delta m$ comes from the radiative
correction, $\delta m\simeq 1/2\alpha_2(m_W-c^2_W m_Z)\sim 0.18$~MeV
for $m \gg m_W$ and $m_Z$. Here, $m_W$ and $m_Z$ are the $W$ and $Z$
boson masses, respectively, and $c_W(\equiv \cos\theta_W)$ is for the
Weinberg angle.  Effective higher-dimensional operators, such as $h^4
T^2/\Lambda^3$, also generate $\delta m$, however they are suppressed by
the new particle mass scale $\Lambda$.  The thermal relic density of the DM
with mass around 1.7~TeV is consistent to the WMAP data.

Another example for the EWIMP DM's is a neutral component of  a pair of
SU(2)$_L$-doublet fermions ($D$ and $D'$) with the
hypercharges $\pm 1/2$. 
This corresponds to  the Higgsino-like LSP in the SUSY models. The
$\chi^0$ is accompanied with a neutral Majorana fermion,
$\chi^{\prime 0}$, in addition to a charged Dirac fermion,
$\chi^{\pm}$. They are again degenerated in mass in the $SU(2)_L$ 
symmetric limit. 
The mass difference is generated by the effective 
operators, such as $ h^2 D^2/\Lambda$, via the EW symmetry breaking.
The thermal relic density of the DM explains  the WMAP data when the
mass is around 0.6~TeV.

In the current Universe the DM is expected to be highly
non-relativistic as mentioned before. In this case, the perturbative
pair annihilation cross sections of the EWIMP DM may have bad
behaviors if the DM mass is heavier than the weak scale. One of the
example is the annihilation cross section to $2\gamma$ at the leading
order.  The process is induced at one-loop level, and the cross
section is $4(1/4)\pi
\alpha^2\alpha_2^2/m_W^2$ for the SU(2)$_L$-triplet (doublet) DM in the SU(2)$_L$
symmetric limit. The cross section is not suppressed by $1/m^2$, and
the perturbative unitarity is violated when $m$ is heavy enough.

The NR effective theory is useful to evaluate the cross sections in the
NR limit. In Ref.~\cite{Hisano:2002fk} we studied the NR effective
theory for the EWIMP in a perturbative way and found that the trouble
in the cross section to $2\gamma$ is related to the threshold
singularity. In order to evaluate the cross section quantitatively,
we have to calculate the cross section  non-perturbatively using the 
NR effective theory \footnote{ The long-distance effect is 
negligible for the DM annihilation in the early universe since the
the velocity at the decoupling temperature is larger than $\alpha_2$.}.  

For evaluation of the annihilation cross sections for heavy EWIMP, we
need to solve the EWIMP wave function under the EW potential. In this
paper, we show the formulae for evaluating the cross sections in the
SU(2)$_L$-triplet DM case. Those for the SU(2)$_L$-doublet case
will be shown in the further publications
\cite{HMN}.

The NR effective Lagrangian for two-body states, $\phi_N({\bf
r})(\simeq 1/2 \chi^0 \chi^0)$ and $\phi_C({\bf r})(\simeq
\chi^-\chi^+)$,  is given as
\begin{equation}
{\cal L}
=
\frac12 {\bf \Phi}^{T}({\bf r})
\left(\left(E+\frac{\nabla^2}{m}\right) {\bf 1}
-{\bf V}(r)+2 i {\bf \Gamma} \delta^3({\bf r})\right)
{\bf \Phi}({\bf r})~, 
\label{Leff}
\end{equation}
where ${\bf \Phi}({\bf r}) = (\phi_C({\bf r}),~\phi_N({\bf r}))$,
${\bf r}$ is the relative coordinate ($r=|{\bf r}|$), and $E$ is the
internal energy of the two-body state.  The EW potential ${\bf V}(r)$
is
\begin{equation}
{\bf V}(r)
=
\left(
\begin{array}{cc}
2 \delta m 
- {\displaystyle\frac{\alpha}{r}}
- {\alpha_2 c_W^2}  {\displaystyle\frac{e^{-m_Zr}}{r}}
&
- {\sqrt{2}\alpha_2}{\displaystyle\frac{e^{-m_Wr}}{r}}
\\
- {\sqrt{2}\alpha_2}{\displaystyle\frac{e^{-m_Wr}}{r}}
&
0
\end{array}
\right).
\end{equation}
In this equation we keep only $2\delta m$ in (1,1) components in order to
calculate the DM annihilation rate up to $O(\sqrt{\delta m/m})$ 
\cite{Hisano:2002fk}. ${\bf \Gamma}$ is the absorptive part of the two-point functions.
Note that a factor of $1/2$ ($1/\sqrt{2}$) is multiplied for ${\bf
V}_{22}$ and ${\bf \Gamma}_{22}$ (${\bf V}_{12}$ and ${\bf
\Gamma}_{12}$) since $\phi_N$ is a two-body state of
identical particles. Thus, ${\bf \Gamma}_{22}$ (${\bf \Gamma}_{11}$) is 
the tree-level annihilation cross section multiplied by the relative
velocity $v$ and $1/2(1)$.  Since the SU(2)$_L$-triplet DM is
assumed to be a Majorana fermion, the $^1S$-wave gauge contribution to
${\bf \Gamma}$ is relevant to the NR annihilation, and then,
\begin{equation}
{\bf \Gamma} =
\frac{\pi\alpha_2^2}{m^2}
\left(
\begin{array}{cc}
\frac{3}2 &\frac1{2\sqrt{2}} \\
\frac1{2\sqrt{2}} & 1
\end{array}
\right)~.
\end{equation}

The annihilation cross section of $\chi^-\chi^+$ or $2\chi^0$ to
the EW gauge boson pair can be expressed using the two-by-two Green
function, ${\bf G}({\bf r},{\bf r^\prime})$, which is given by
\begin{eqnarray}
&\left(\left(E+{\displaystyle\frac{\nabla^2}{m}}\right) {\bf 1}-{\bf V}(r)+2 i {\bf \Gamma}
\delta^3({\bf r})\right){\bf G}({\bf r},{\bf r^\prime})&
\nonumber\\
&= \delta^3({\bf r}-{\bf r^\prime})
{\bf 1}~.&
\label{eqgr}
\end{eqnarray}
Due to the optical theorem, the long-distance (wave function) and the
short-distance (annihilation) effects can be factorized
\cite{Bodwin:1994jh}. The annihilation cross sections to $VV^\prime$
($V,V^\prime= W,Z,\gamma$) are written as
\begin{eqnarray}
&(\sigma v)_{VV^\prime} = c_i \sum_{ab} {\bf \Gamma}_{ab}|_{VV^\prime}& \times
A_a A_b^\star~,
\end{eqnarray}
where $i$ represents the initial state ($i=0$ and $\pm$ for $2\chi^0$
and $\chi^-\chi^+$ pair annihilation, respectively) and $A_a =\int d^3
r~ {\rm e}^{-i {\bf k} {\bf r}} (E +{\nabla^2}/{M}) {\bf G}_{ai}({\bf
r},0)$ with $k=\sqrt{m E}=mv/2$.  Here $c_0=2$ and $c_{\pm}=1$, where
$c_0$ is a factor needed to compensate the symmetric factor for ${\bf
\Gamma}$ and ${\bf V}$.  ${\bf
\Gamma}_{ab}|_{VV^\prime}$ is the contribution to ${\bf \Gamma}_{ab}$
from the final states $VV^\prime$. It is clear that if the
long-distance effect is negligible, $(\sigma v)_{VV^\prime}= c_i {\bf
\Gamma}_{ii}|_{VV^\prime}$.

The $S$-wave annihilation is dominant in the NR annihilation.  Thus, the
Green function is reduced to ${\bf G}({\bf r},{\bf r^\prime}) ={\bf
g}(r,r^\prime)/rr^\prime$.  Similar to the case in one-flavor system,
we find that ${\bf g}(r,r^\prime)/rr^\prime$ is expressed by the independent solutions of
the homogeneous part of the Eq.~(\ref{eqgr}), ${\bf g}_>(r)/r$ and ${\bf
g}_<(r)/r$, as
\begin{eqnarray}
{\bf g}(r,r^\prime)
&=&
\frac{m}{4\pi} 
{\bf g}_>(r) {\bf g}_<^T(r^\prime)\theta(r-r^\prime)
\nonumber\\
&&+
\frac{m}{4\pi} 
{\bf g}_<(r) {\bf g}_>^T(r^\prime)\theta(r^\prime-r)~.
\end{eqnarray}
The solutions ${\bf g}_>(r)$ and ${\bf g}_<(r)$ are also two-by-two
matrices since $\Phi({\bf r})$ has two degrees of freedom.  The
boundary conditions at $r=0$ are ${\bf g}_<(r)|_{r\rightarrow 0} = {\bf 0}$,
${\bf g}_<^\prime(r)|_{r\rightarrow 0} = \bf{1}$, and $ {\bf g}_>(r)
|_{r\rightarrow 0} = {\bf 1}$.  In the following, we assume $E<2\delta
m$ so that a pair annihilation of $\chi^0$ does not produce on-shell
$\chi^-\chi^+$. As the result,
\begin{equation}
{\bf g}_>(r) |_{r\rightarrow \infty} =
\left(\begin{array}{cc}
0&0\\
d_{1} {\rm e}^{ik r}& d_{2} {\rm e}^{ik r}
\end{array}
\right)~.
\end{equation}
In this case, the $\chi^0$-pair annihilation cross sections are
$(\sigma v)_{VV^\prime}= c_i \sum_{ab} {\bf \Gamma}_{ab}|_{VV^\prime}
d_{a} d_{b}^{\star}$, as expected.  It is enough to calculate $d$ in
order to evaluate the cross sections.

\begin{figure}[t]
\begin{minipage}{0.45\linewidth}
{\psfig{file=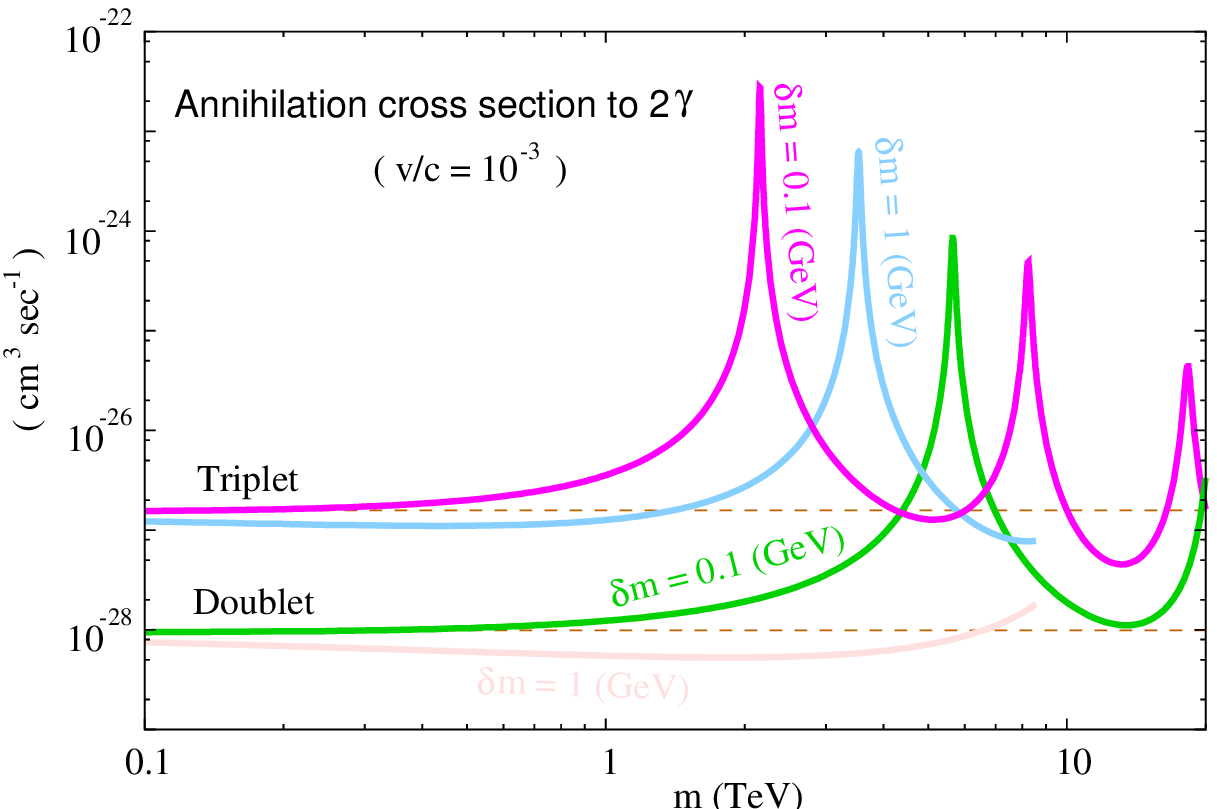,width=3.4in,angle=0}}
\end{minipage}
\begin{minipage}{0.45\linewidth}
{\psfig{file=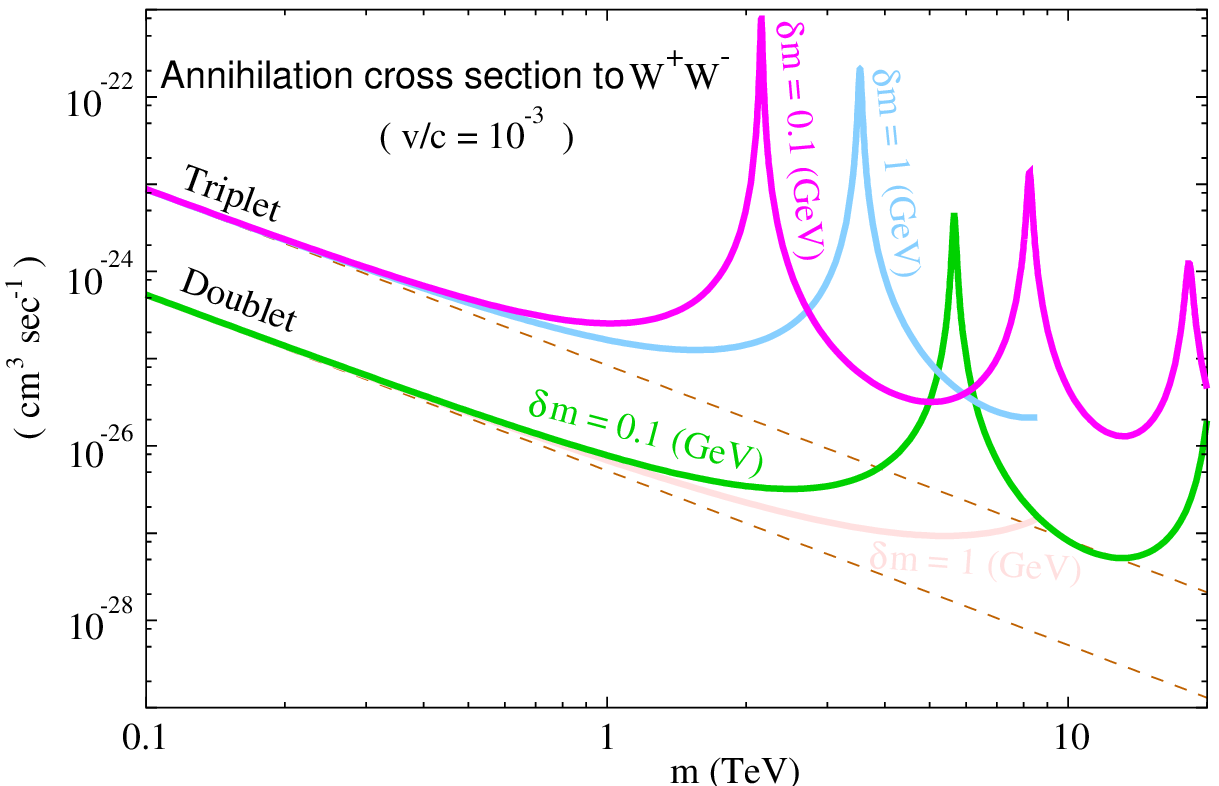,width=3.4in,angle=0}}
\end{minipage}
\caption{The $\chi^0$-pair annihilation cross sections 
to $2\gamma$ and $W^+W^-$ when $\delta m=0.1$~GeV (slid lines) and
1~GeV (dashed lines).  $\chi^0$ is the SU(2)$_L$-triplet or doublet
DM. Here, $v/c=10^{-3}$.  The leading-order cross sections in the perturbation
are also shown for $\delta m=0$ (dotted lines).
}
\label{fig1}
\end{figure}

In Fig.~(\ref{fig1}) we show the annihilation cross sections of the
SU(2)$_L$-triplet DM pair to $2\gamma$ and $W^+W^-$ as functions of
$m$. We evaluated the cross section numerically.  Here, we take
$v/c=10^{-3}$, which is the typical averaged velocity of the DM in our
galaxy, and $\delta m=0.1$~GeV and 1~GeV. The perturbative cross
sections are  also plotted. Large $\delta m$ leads
to unreliable numerical calculation for large $m$, and
then some curves are terminated at some points. However, $\delta m$
should be suppressed around the regions.

When $m$ is around 100~GeV, the cross sections to $2 \gamma$ and $W^+W^-$
are almost the same as the perturbative ones.
The cross section to $2\gamma$ is suppressed by a loop factor there.
However, when $m\gsim 0.5$~TeV, the cross sections are significantly
enhanced and have the resonance structure. Especially, the cross
section to $2\gamma$ becomes comparable to that to $W^+W^-$ around the
resonance. This suggests that the $2\chi^0$ state is strongly mixed
with $\chi^+\chi^-$.

\onecolumngrid

\twocolumngrid
The qualitative behavior of the cross sections around the first
resonance may be understood by approximating the EW potential by a
well potential. Taking $c_W=1$ for simplicity, the EW potential is
approximated as
\begin{equation}
{\bf V}(r)
=
\left(
\begin{array}{cc}
2 \delta m 
- b_1{\alpha_2 }  m_W
&
- b_1{\sqrt{2}\alpha_2} m_W
\\
- b_1{\sqrt{2}\alpha_2} m_W
&
0
\end{array}
\right)~,
\end{equation}
for $r<R(\equiv ({b_2} m_W)^{-1})$. Here, $b_1$ and $b_2$ are
numerical constants. By comparing the annihilation cross sections to $2
\gamma$ in this potential and in the perturbative calculation for
small $m$, we find $b_1=8/9$ and $b_2=2/3$. Under this potential,
two-body states $2\chi^0$ and $\chi^-\chi^+$ have the attractive and
repulsive states, whose potential energies are $\lambda_{\pm}=1/2
({\bf V}_{11}\pm\sqrt{{\bf V}^2_{11}+4{\bf V}^2_{12}})$ with ${\bf
V}_{ij}(i,j=1,2)$ elements in ${\bf V}$. The attractive state is
$-\sin\theta
\phi_C+\cos\theta \phi_N$ with $\tan^2\theta=\lambda_-/\lambda_+$. 

When $\delta m\ll b_1\alpha_2 m_W/2(\sim 1$~GeV), $\theta$ is not
suppressed by $\delta m$ and $\chi^-\chi^+$ and $2\chi^0$ are
mixed under the potential. In this case, the cross section to
$2\gamma$ is given as
\begin{eqnarray}
(\sigma v)_{2\gamma}
&=&
\frac{4\pi\alpha^2}{9 m^2}\left(
\frac1{\cos(k_-R)}-\frac1{\cosh(k_+R)}\right)^2~,
\label{appro}
\end{eqnarray}
where $k_\pm^2 = |\lambda_\pm| m$.  Here, we neglect the ${\bf \Gamma}$
term contribution to the wave function for simplicity and take
$E\simeq 0$. The cross section (\ref{appro}) is reduced to
$4\pi\alpha^2\alpha_2^2/m_W^2$ for ${\alpha_2 m}\lsim m_W$. On the other
hand, it is not suppressed by a one-loop factor for ${\alpha_2 m}\gsim
m_W$ and has a correct behavior as $\sim 1/m^2$ in a heavy $m$ limit.
When $k_-R=(2n-1)\pi/2$ ($n=1,2,\cdots$), the zero energy resonance,
whose binding energy is zero, appears and the cross section is
enhanced significantly. In Fig.~(\ref{fig1}), the $n$-th zero energy
resonance  appears at $m=m^{(n)}\sim n^2\times m^{(1)}$,
while the well potential predicts $m^{(n)}\sim (2n-1)^2
\times m^{(1)}$. We guess that the Yukawa potential might be
approximated better by the Coulomb potential for the higher zero
energy resonances.

When the zero energy resonance exits, the cross sections $\sigma v$
are proportional to $v^{-2}$ for $v\ll 1$. However, this is not a signature for
breakdown of the unitarity. We find from study in the one-flavor
system under the well potential $V$ that when $v \ll m V \Gamma$,
$\sigma v$ is saturated by the finite width $\Gamma$ and the unitarity
is not broken.

We also show the annihilation cross sections for the SU(2)$_L$-doublet
DM in Fig.~(\ref{fig1}). The SU(2)$_L$-doublet DM has the smaller
gauge charges compared with the SU(2)$_L$-triplet DM. As the result,
the cross section is smaller, and the first zero energy resonance
appears at 5~TeV.

The enhancement for the DM annihilation rates gives significant
impacts on the indirect searches for the DM in the cosmic ray. In the
following we discuss the $\gamma$ ray search from the DM annihilation
in the galactic center and the future prospects.

The line $\gamma$ from the pair annihilation to
$2\gamma$ or $Z\gamma$ at the galactic center is a robust signal for
the DM. Also, $W^-W^+$ and $2 Z^0$ final states produce the continuum
$\gamma$ spectrum through $\pi^0\rightarrow 2\gamma$, and the
observation may constrain the EWIMP DM.  The $\gamma$ flux,
$\Psi_\gamma(E)$, is given as
\begin{eqnarray}
&{\displaystyle\frac{d\Psi_\gamma(E)}{d E}} = 
9.3 \times 10^{-12} {\rm cm}^{-2} {\rm
sec}^{-1}{\rm GeV^{-1}} \times \bar{J} \Delta \Omega&
\nonumber\\
&\times\left(
{\displaystyle\frac{100{\rm GeV}}{m}}
\right)^2 
{\displaystyle\sum_{VV^\prime}} 
{\displaystyle\frac{dN^{{VV^\prime}}}{dE}}
\left(
{\displaystyle
\frac{\left<\sigma v\right>_{VV^\prime}}
{10^{-27} {\rm cm^3 sec^{-1}}}
}\right)~.&
\end{eqnarray}
where $N^{{VV^\prime}}$ is the number of photons from the final state
$VV^\prime$ and $\left<\sigma v\right>$ is the averaged cross section
by the velocity distribution function.  Note that the angular
acceptance of the detector, $\Delta\Omega$, is $10^{-3}$ typically for
the ACT detectors. The flux depends on the halo DM density profile
$\rho$ through
\begin{equation}
\bar{J} \Delta \Omega  = \frac{1}{8.5{\rm kpc}}\int_{\begin{array}{l}
\mbox{l.o.s}\\
\Delta \Omega
\end{array}
} d\Omega dl~
\left(\frac{\rho}{0.3~{\rm GeV cm^{-3}}}\right)^2~,
\end{equation}
where the integral is along the line of sight. 
$\bar{J}$ is studied for various
halo models, and $3\lsim \bar{J} \lsim 10^5$
\cite{Bergstrom:1997fj}. The cuspy structures in the halo density profiles,
which are suggested by the $N$-body simulations, tend to give larger numbers
to $\bar{J}$. In the following we take a moderate value as $\bar{J}=500$,
which is typical  for the NFW profile \cite{nfw}.

\onecolumngrid

\begin{figure}[t]
\begin{minipage}{0.45\linewidth}
{\psfig{file=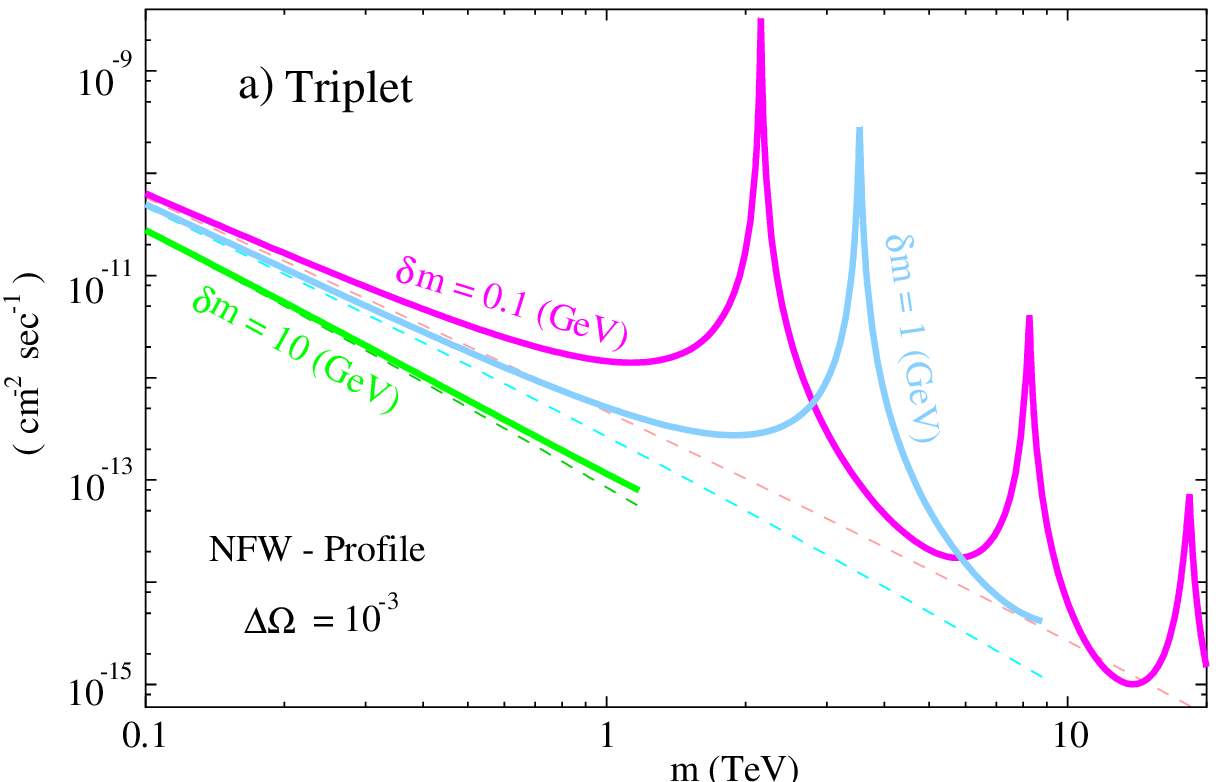,width=3.4in,angle=0}}
\end{minipage}
\begin{minipage}{0.45\linewidth}
{\psfig{file=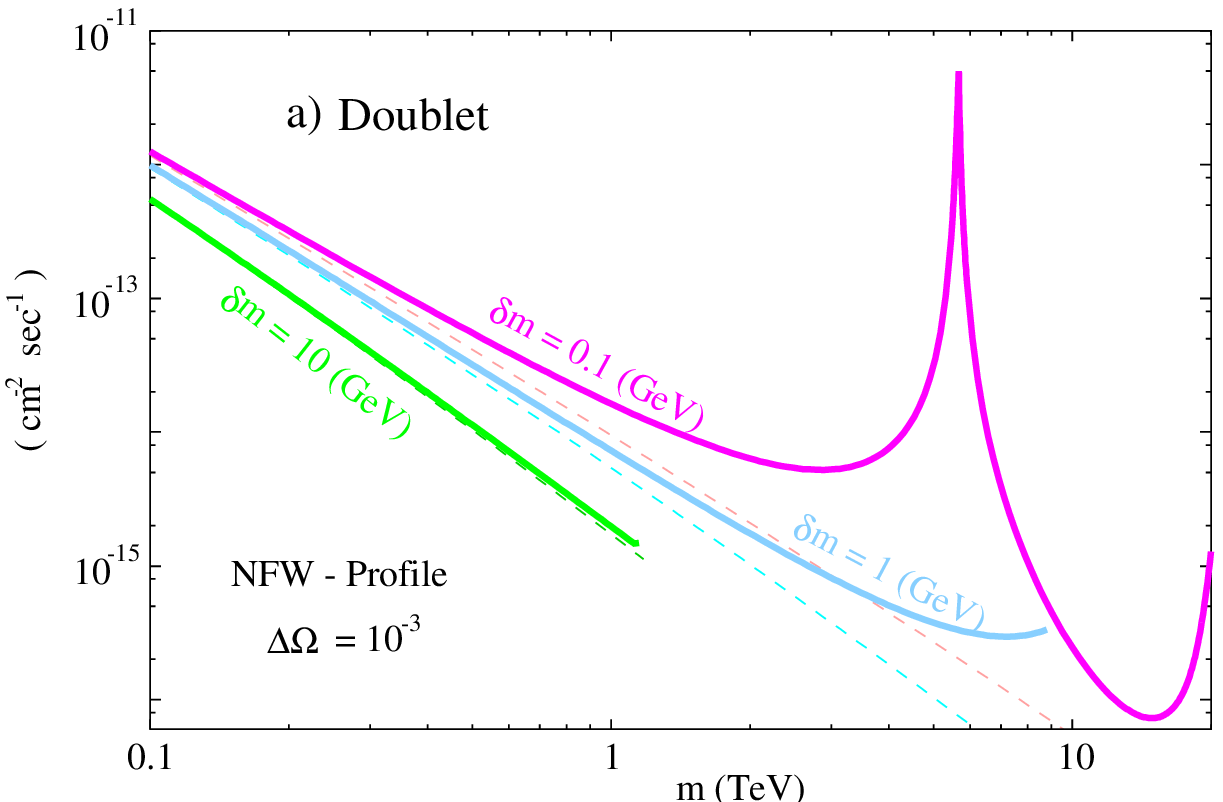,width=3.4in,angle=0}}
\end{minipage}
\begin{minipage}{0.45\linewidth}
{\psfig{file=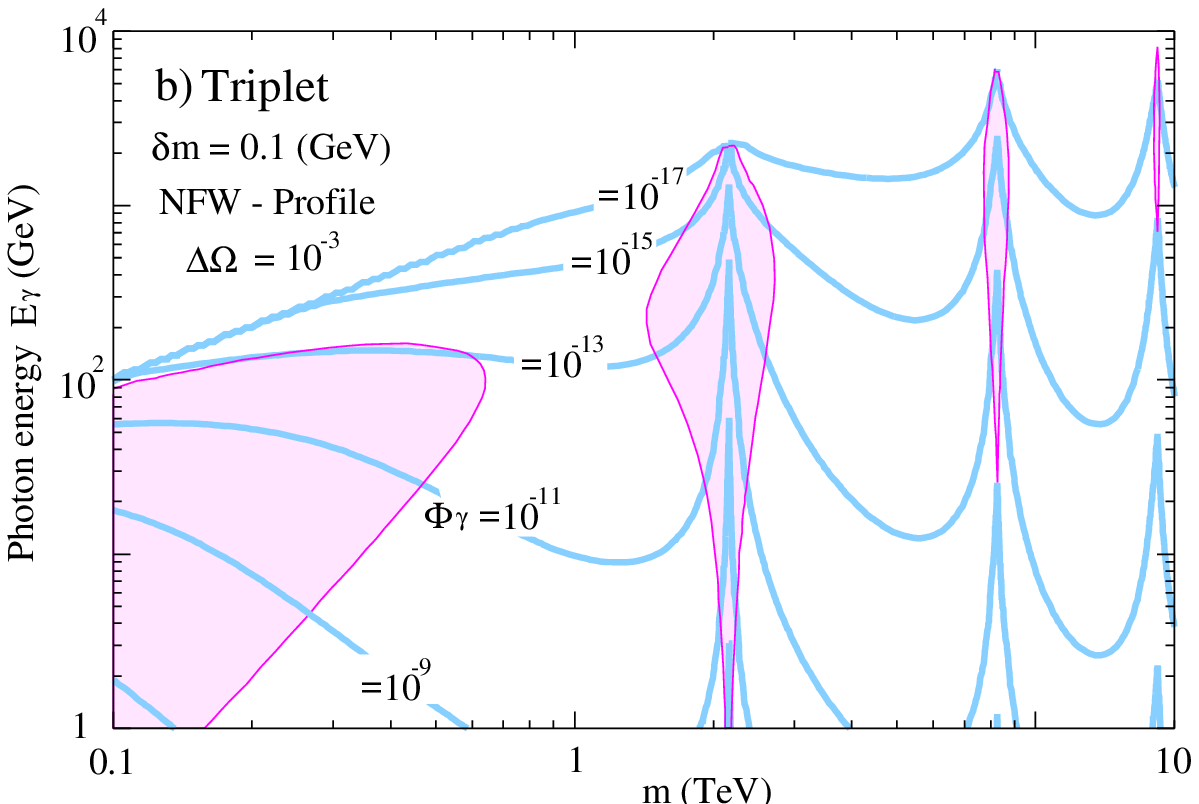,width=3.4in,angle=0}}
\end{minipage}
\begin{minipage}{0.45\linewidth}
{\psfig{file=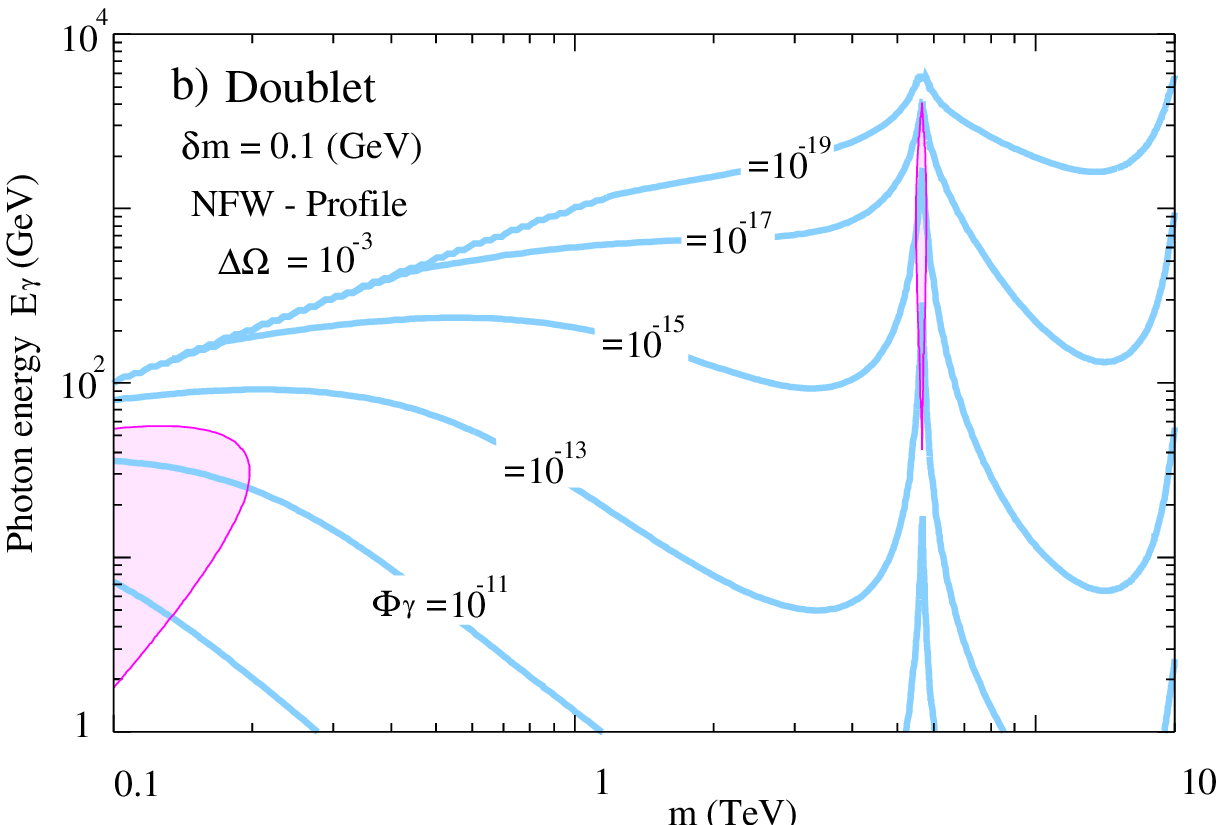,width=3.4in,angle=0}}
\end{minipage}
\caption{a) Line $\gamma$ flux  from the galactic center 
in cases of the SU(2)$_L$-triplet and doublet DM's.  We take the
averaged velocity of the DM $v/c=10^{-3}$, $\bar{J}=500$ and $\Delta
\Omega=10^{-3}$.  The leading-order cross sections in the perturbation
are also shown (dotted lines). b) Contour plot of continuum $\gamma$
ray flux from the galactic center in the unit of
cm$^{-2}$sec$^{-1}$GeV$^{-1}$. Here, we take $\delta m=0.1$~GeV, and
other parameters are same as in a).  The shaded regions correspond to
$S/B>1$.}
\label{fig2}
\end{figure}

\twocolumngrid
In Fig.~(\ref{fig2}a) we show the line $\gamma$ flux from the galactic
center in cases of the SU(2)$_L$-triplet and doublet DM's. Here, we
take $\delta m=0.1,1,10$~GeV. We also show the flux obtained by the
leading-order calculation for comparison.  The ACT detectors have high
sensitivity for the TeV-scale $\gamma$ ray. MAGIC and VERITAS in the
north hemisphere might reach to $10^{-14}$cm$^{-2}$s$^{-1}$ at the TeV
scale while CANGAROO III and HESS in the south hemisphere to
$10^{-13}$cm$^{-2}$s$^{-1}$ \cite{Bergstrom:1997fj}.  These ACT
detectors are expected to cover the broad region.

In Fig.~(\ref{fig2}b), the contour plot of the continuum $\gamma$ flux
from the galactic center is presented. For ${dN^{{VV^\prime}}}/{dE}$
we use the fitting functions given in \cite{Bergstrom:1997fj}. Shaded
regions correspond to $S/B>1$.  In order to evaluate the background
$B$, we assume a power low fall-off in the energy for the diffused
$\gamma$ ray flux $\Psi_{BG}(E)$ as ${d\Psi_{BG}(E)}/{dE} =9.1\times
10^{-5} {\rm cm^{-2} sec^{-1} GeV^{-1}} \times ({{E}/{\rm
1~GeV}})^{-2.7} \Delta \Omega$ \cite{Bergstrom:1997fj}.  The EGRET
experiment has observed the diffused $\gamma$ ray emission from the
galactic center up to about 10~GeV \cite{egret}.  Even the small
regions around the resonances in addition to the triplet DM with
$m_{\chi}=100$~GeV are already constrained by the EGRET observation.
The GLAST satellite detector, which will detect $\gamma$ ray with
1~GeV$<E<$300~GeV, have more sensitivity to the around the region.

\underline{Acknowledgment}\\
This work is supported in part by the Grant-in-Aid for Science
Research, Ministry of Education, Science and Culture, Japan
(No.15540255, No.13135207 and No.14046225 for JH and No.14540260 and
14046210 for MMN).

\end{document}